\documentclass{article}
\usepackage[T1]{fontenc}
\usepackage[utf8]{inputenc}
\usepackage{ismir} 
\usepackage{amsmath,cite,url}
\usepackage{graphicx}
\usepackage{color}
\usepackage{multirow}
\usepackage{booktabs}
\usepackage{xcolor}
\usepackage{microtype}

\newcommand{\ours}{RNHybrid }

\title{From Prediction to Collaboration: \\ Interactive Symbolic Music Analysis}

\multauthor
  {Emmanouil Karystinaios$^1$ \hspace{0.3cm} Johannes Hentschel$^2$ 
  \hspace{0.3cm} Markus Neuwirth$^2$ \hspace{0.3cm} Gerhard Widmer$^1$
  }
  {
  $^1$ Department of Computational Percepetion, Johannes Kepler University Linz, Austria\\
  $^2$ Anton Brukner University Linz, Austria\\  
  {\tt\small firstname.lastname@jku.at}
  }

\def\authorname{E. Karystinaios, J. Hentschel, M. Neuwirth, and G. Widmer}

\begin{document}

\maketitle

\begin{abstract}
Automatic symbolic music analysis has made substantial progress, yet existing systems are typically designed for a single mode of use, such as full-score prediction, and therefore do not match the broader range of operations that arise in analysis workflows, including partial completion, local correction, and iterative refinement. As a result, there remains a gap between strong benchmark models and systems that can support interactive analytical use. We present a unified framework for symbolic Roman-numeral (RN) analysis that narrows this gap by combining strong predictive performance with direct support for constrained completion and revision. The method is designed to provide a practical trade-off between accuracy and interactive responsiveness by computing expensive pretrained representations once and reusing them during iterative refinement, making powerful pretrained models more amenable to interactive settings. It supports complete score analysis, targeted revision of existing labels, and inference of missing annotations from partial context through a shared modeling framework. Experiments on Dilemmadata, the largest and most heterogeneous benchmark of its kind, show that the proposed approach is a strong RN-analysis baseline while also supporting masked completion from partial labels. Together with a prototype interface for multi-level candidate inspection and editing, these results position automatic RN analysis not only as a prediction problem, but also as a foundation for future interactive tools for music analysis.
\end{abstract}

\section{Introduction}\label{sec:introduction}

Roman numeral (RN) analysis describes chords and harmonic functions relative to a local key. It is widely used for Western tonal music and can also be applied to some modal and extended-tonal repertoire. Automatic RN analysis remains challenging because harmonic labels are highly context-dependent and must stay coherent across multiple musical layers. While recent systems have improved benchmark accuracy, they are still mostly framed as one-shot full-score predictors. This evaluation setting only partially matches musicological practice, where analysts iteratively inspect uncertain passages, correct local mistakes, and complete missing labels from partial context. As a result, there is still a practical gap between strong RN predictors and tools that genuinely support analyst-in-the-loop workflows.

Ongoing research has progressed along two complementary directions. Pretrained symbolic encoders, especially MusicBERT and its RN-focused adaptation RNBert, provide rich contextual representations and strong RN performance \cite{zeng2021musicbert,sailor2024rnbert}. In parallel, graph-based approaches such as ChordGNN and AnalysisGNN model explicit relational structure and functional dependencies across musical events \cite{karystinaios2023chordgnn,kernfeld2025analysisgnn}. These strengths are naturally synergistic: sequence pretraining captures broad musical context, while graph reasoning supports coherent decisions across note-, onset-, and beat-level structures. Motivated by this complementarity, we build a hybrid backbone that couples MusicBERT embeddings with an AnalysisGNN-style graph module, and use it as the prediction engine for an analyst-in-the-loop workflow that spans blind RN analysis, partial-label completion, and interactive revision.

Our model integrates information at multiple granularities (note, onset, and beat) through a shared graph representation. Fine-grained prediction is performed at the note level, including prediction of note-level functional roles. For coarser musical decisions, we introduce post-hoc voters at onset and beat levels that aggregate selected tasks such as full Roman numeral labels, cadence, and phrase-related structure. This design preserves local detail while enabling higher-level harmonic decisions that are useful for full-score inference and for downstream editing workflows.

We evaluate on Dilemmadata~\cite{Hentschel2026_DilemmadataSymbolicDataset}, a new RN analysis benchmark that combines the AugmentedNet dataset~\cite{NapolesLopez2021_AugmentedNetRomanNumeral} and the Distant Listening corpus~\cite{hentschel2025distantlistening} into a larger and more heterogeneous evaluation setting. In addition to blind full-score analysis, we study masked prediction for partial-label completion and iterative correction. Together, these capabilities position RN analysis not only as a prediction task, but also as a setting for studying model behavior under human edits and partial analytical context.

\begin{figure*}[!ht]
    \centering
    \includegraphics[width=\linewidth]{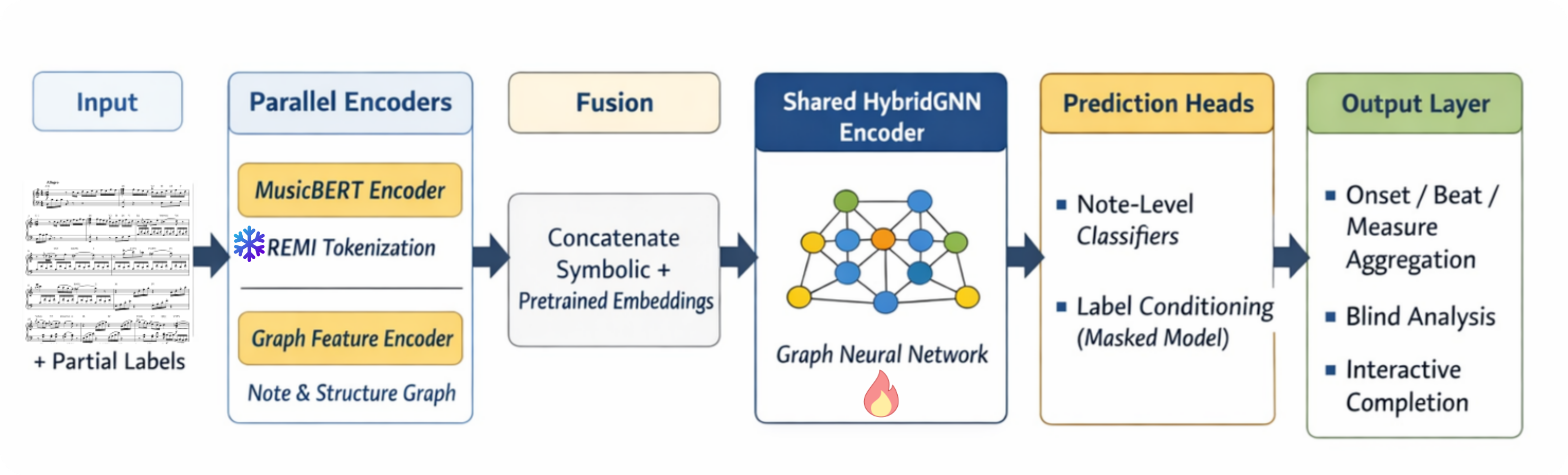}
    \caption{Overview of the proposed architecture. A MusicBERT-based sequence branch produces contextual note representations, which are fused with symbolic note features and processed by a graph module. Task-specific heads produce note-level predictions, which can then be aggregated to onset- and beat-level decisions for analysis and interactive editing.}
    \label{fig:architecture}
\end{figure*}  

In summary, this paper makes four contributions. First, we recast automatic RN analysis as a prediction-to-analysis continuum in which blind full-score prediction, partial-label completion, and constrained revision can be addressed within a single framework. Second, we introduce an edit-conditioned masked-prediction model that supports partial-label completion and iterative revision with hard constraints on known labels. Third, we develop a prototype interface that turns per-note probabilities into top-$k$ Roman-numeral candidates at note, onset, beat, and measure level, enabling inspection and editing across multiple musical scopes. Fourth, we evaluate the framework on Dilemmadata, where \ours yields competitive blind inference and the masked checkpoint improves monotonically as more contextual labels are provided.\footnote{Interactive interface: \url{analysisgnn.com} \\ 
Codebase:  \url{https://github.com/manoskary/analysisgnn}}

\section{Related Work}\label{sec:related_work}

Prior work on automatic Roman numeral analysis has largely focused on improving blind full-score prediction. Early systems used recurrent and multi-task formulations \cite{chen2018functional,micchi2020notallroads}, while later attention-based models improved the handling of longer-range harmonic context \cite{chen2021attend}. AugmentedNet further strengthened this line of work by combining multi-task tonal prediction with synthetic data augmentation across multiple corpora, showing that redundant analysis tasks improve prediction~\cite{NapolesLopez2021_AugmentedNetRomanNumeral}. These systems establish strong automatic baselines, but they are primarily evaluated as one-shot predictors rather than as tools for iterative analytical workflows.

A second line of work focuses on harmonic coherence and structured output dependencies. Micchi \textit{et al.} introduced \textit{frog}, which uses an autoregressive output layer to enforce consistency among harmonic sub-labels \cite{micchi2021frog}. Their formulation is based on Neural Autoregressive Distribution Estimation (NADE) \cite{larochelle2011nade}, adapted to symbolic harmony modeling. This emphasis on label consistency is directly relevant to our setting, because interactive correction and partial completion require not only local accuracy but also compatibility among related harmonic decisions.

Recent work has also expanded the representational and architectural toolkit available for RN analysis. ChordGNN models score structure directly at the note level and predicts onset-wise harmonic labels via graph neural networks \cite{karystinaios2023chordgnn}, while AnalysisGNN extends this direction toward unified multi-task music analysis (20 distinct properties for each note) with graph-based architectures~\cite{kernfeld2025analysisgnn}. In parallel, self-supervised pretraining has become increasingly influential: MusicBERT provides strong symbolic representations \cite{zeng2021musicbert}, and RNBert shows that fine-tuning these representations can substantially improve RN-analysis accuracy \cite{sailor2024rnbert}. A complementary recent direction targets musicological applications more directly by introducing an interpretable RNBERT variant based on multilinear mixture-of-experts \cite{triantafyllou2026interpretable}. Our work builds directly on these two strands by combining pretrained symbolic encoders with graph-based relational reasoning in a single framework.

Finally, broader and more diverse corpora have enabled stronger and more realistic evaluation settings, including When in Rome \cite{gotham2023wheninrome} and the Distant Listening Corpus \cite{hentschel2025distantlistening}. We build on this progress while targeting a use case that remains less emphasized in recent RN work: a single system that supports blind prediction, constrained completion from partial labels, and iterative analyst-in-the-loop revision.

\section{Methodology}\label{sec:methodology}

\begin{figure*}[ht]
    \centering
    \includegraphics[width=0.8\linewidth]{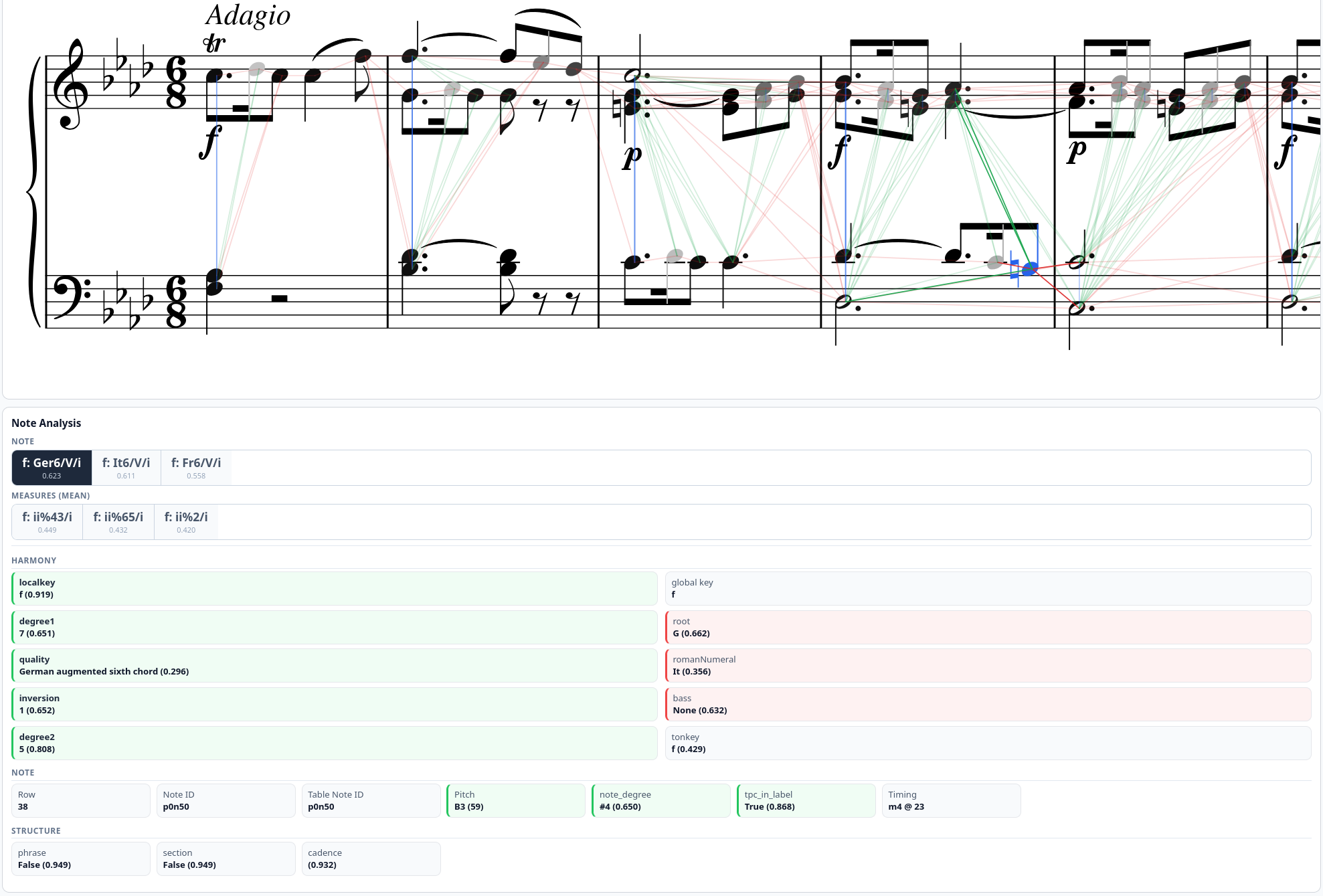}
    \caption{Interactive interface (Verovio). Clicking a note (here the B at the end of m.~4, left hand) surfaces the top-3 Roman-numeral candidates for that note and for its enclosing onset, beat, and measure spans (measure top-3 shown). Selecting a candidate, here ``f: Ger6/V/i'', recolours the task-level predictions in green (agree) or red (contradict); three tasks contradict the selection in this example. Edits in the task table feed back into the masked-completion model to regenerate dependent labels.}
    \label{fig:interface}
\end{figure*}

Our goal is to support three complementary analysis modes within one framework: (i) complete-score inference, (ii) local revision of existing labels, and (iii) completion of partially annotated passages. Our system combines an RNBert-inspired sequence component based on pretrained representations with an AnalysisGNN-style multi-task graph model \cite{sailor2024rnbert,kernfeld2025analysisgnn}. 

\subsection{Problem Formulation}

Given a symbolic score $S$, we predict note-level labels for Roman-numeral components and related structural tasks. Following common RN decompositions \cite{NapolesLopez2021_AugmentedNetRomanNumeral,Hentschel2021_AnnotatedMozartSonatas,karystinaios2023chordgnn,sailor2024rnbert}, we frame harmonic analysis as multi-task classification over local key, scale degree, inversion, chord quality, complete Roman numeral, and auxiliary labels such as cadence, phrase, section, pedal-points, root, bass, functional degree of every note, and non-chord-tone status, with a total of 20 analysis tasks. Note-level predictions are the main model output; onset-, beat-, and potentially measure-level reductions are computed afterwards by aggregation modules. 

\subsection{Complete Inference Model}

The complete-analysis model is a multi-task graph neural network over symbolic note graphs, augmented with note representations from our REMI-tokenized MusicBERT variant~\cite{karystinaios2026musicbertlarge}. In the sequence branch, each score is tokenized with a REMI-style vocabulary and aligned to notes through precomputed token-to-note maps. For long scores, the token sequence is divided into chunks matching the model's maximum positional length, and the resulting contextual states are pooled to note level.

The graph branch operates on note-level features, including pitch spelling and key-signature embeddings, together with edges encoding simultaneity, temporal succession, overlap, and rest relations~\cite{Karystinaios2024_GraphMuseLibrarySymbolic}. The model instantiates beat and measure nodes, which support hierarchical aggregation and prediction at multiple musical levels. MusicBERT note embeddings are concatenated with the symbolic note features before graph encoding, and decoded by task-specific classification heads that predict RN components along with other note-level tasks such as cadence, phrase, section, root, bass, and non-chord-tone labels.

\subsection{Masked Edit-Conditioned Model}

Interactive completion and revision are handled by a model specialized for masked prediction. This model is distilled from the full-inference model and extended with label-conditioning modules. For a selected subset of tasks, known labels are embedded and injected into the encoder as a note-level conditioning bias before message passing. Unknown positions receive a masked embedding, while known labels are enforced through hard constraints at inference time.

Training uses explicit masked task selection together with a hybrid masking policy that combines random node masking and onset-span masking. Supervision for masked tasks is concentrated on target nodes, while known context nodes act as conditioning evidence. This design makes the masked checkpoint suitable for partial completion and analyst-in-the-loop revision rather than as a direct replacement for blind full-score inference.

\begin{table*}[ht]
\centering
\begin{tabular}{c|l|c|c|c|c|c|c|c|c|c}
\textbf{Dataset} & \textbf{Model} & \textbf{Degree} & \textbf{Quality} & \textbf{Inversion} & \textbf{Local Key} & \textbf{Roman} & \textbf{Cadence} & \textbf{Phrase} \\
\hline
\multirow{6}{*}{AugNet} 
    & AugmentedNet~\cite{NapolesLopez2021_AugmentedNetRomanNumeral}            & .670 & .797 & .788 & .829 & .464          &  --  &  -- \\ 
    & ChordGNN+Post~\cite{karystinaios2023chordgnn}           & .714 & .784 & \textbf{.803} & .813 & .518          &  --  &  -- \\ 
    & RNBert~\cite{sailor2024rnbert}                  & .731 & \textbf{.819} & .796 & .825 & .574          &  --  &  -- \\     
    & AnalysisGNN~\cite{kernfeld2025analysisgnn}             &  .711  & .801 & .790 & .831 & .530          &  --  &  -- \\     
    & RNBert-style (retrain)  & .683 & .780 & .769 & .799 & .501 &  --  &  -- \\     
    & \ours                  & \textbf{.732} & .815 & .799 & \textbf{.846} & \textbf{.576} &  --  &  -- \\     
\hline\hline
\multirow{4}{*}{DLC} 
    & RNBert~\cite{sailor2024rnbert}                  &  --  &  --  &  --  &  --  &   .301        &      &  --  \\ 
    & AnalysisGNN~\cite{kernfeld2025analysisgnn}             & .680 & .793 & .797 & .828 &   .516        & \textbf{.558} & .742 \\ 
    & RNBert-style (retrain)  & .666 & .777 & .785 & .837 & .509 & .469 & .684 \\ 
    & \ours                  & \textbf{.707} & \textbf{.816} & \textbf{.817} & \textbf{.872} & \textbf{.578} & .529 & \textbf{.755} \\ 
\end{tabular}
\caption{Overview of models by target task. For each dataset, we compare baseline models evaluated on their respective test sets. Cadence and phrase are reported as note-level macro F1 scores; degree, quality, inversion, and local key are reported as accuracies; Roman numerals are evaluated with the CSR score~\cite{NapolesLopez2021_AugmentedNetRomanNumeral}. A Roman numeral is considered correct only when its local key, degree, quality, and inversion are all predicted accurately.}
\label{tab:model_overview}
\end{table*}

\subsection{Optimization and Training}

The base training objective is multi-task cross-entropy with label smoothing across all supervised tasks, combined with feature regularization. For the masked model, a frozen teacher constructed from the full-inference model supplies preservation losses: (i) KL distillation on preserved tasks or non-target nodes, (ii) feature anchoring via mean-squared error between student and teacher note representations, and (iii) L2-SP regularization to penalize drift from initialization. The training loop mixes masked and unmasked batches so that the model learns completion behavior without collapsing full-analysis performance.

Because joint training over ${\sim}20$ note-level tasks exposes strong gradient conflicts, Table~\ref{tab:hybrid_conflict_lora} compares PCGrad~\cite{yu2020pcgrad}, CAGrad~\cite{liu2021cagrad}, and GradNorm~\cite{chen2018gradnorm} as conflict-aware optimizers in the frozen-backbone setting. We do not include NADE-style autoregressive output modeling~\cite{larochelle2011nade,micchi2021frog} in this comparison, as it addresses output-space coherence rather than gradient conflict and has not previously shown clear benefit in graph-based RN work~\cite{karystinaios2023chordgnn}. The strongest runs use AdamW, mixed precision, cosine warmup scheduling, early stopping on validation loss, and transposition augmentation.

\subsection{Aggregation and Post-hoc Decoding}

Note-level predictions are the base output of the model; onset-, beat-, and measure-level analyses are obtained afterwards through post-hoc aggregation. The baseline reduction is simple mean pooling, but the framework also supports learned post-hoc voters that assign note-specific weights before aggregating probabilities, and onset-level beam-search decoders that combine task-wise emission scores with legality checks on reconstructed Roman numerals and transition penalties for abrupt changes in key, degree, or inversion. Under blind full inference, these inference-time modules do not improve accuracy (Table~\ref{tab:ablation_results}); we return to their intended role in Section~\ref{sec:discussion}.

\section{Experiments}\label{sec:experiments}

We evaluate four regimes. First, we measure full inference performance compared to previous models and different model settings. Second, we evaluate masked completion with our edit-conditioned model, where part of the analysis is known and the model predicts the missing labels with explicit known-label preservation. Third, we evaluate different optimization and finetuning scenarios for boosting performance. Finally, we study post-hoc decoding configurations, including onset/beat voters, iterative refinement, and onset-level beam search.

These regimes are not directly interchangeable. Full inference is blind prediction over the whole score, whereas masked completion uses known contextual labels and is therefore a different evaluation problem. We consequently report masked-model metrics separately and do not compare them as if they were full-inference numbers from the same setting.

One recurring issue in this setting is the need to smooth contradictory predictions across related tasks. Our model predicts around 20 labels per note, many of which are partially redundant or are more naturally defined at higher temporal levels such as onsets or segments. As a result, notes within the same onset can occasionally receive inconsistent labels. To address this, we investigate several aggregation strategies, including mean pooling and learned voting. Temporal smoothness is another challenge: because predictions are made at fine granularity, stationary spans can exhibit spurious local label changes. We therefore also evaluate iterative masked prediction with confidence-based locking, beam search, and related post-hoc strategies.

\subsection{Corpora}

All experiments are conducted on Dilemmadata~\cite{Hentschel2026_DilemmadataSymbolicDataset}, a newly proposed benchmark for symbolic music analysis that includes Roman numeral, cadence, phrase, and related structural annotations. 
Within this benchmark, we report results on two test subsets derived from AugmentedNet~\cite{NapolesLopez2021_AugmentedNetRomanNumeral} and the Distant Listening Corpus~\cite{hentschel2025distantlistening}, both processed to share the same analysis vocabulary and task schema. Final numbers are reported only on the held-out AugNet and DLC test subsets, curated to avoid data-leakage. This harmonized preprocessing is important for comparability: both the full-inference and masked-completion experiments use the same shared task vocabulary, even when the underlying source corpora differ in coverage and annotation density.


\subsection{Metrics}

We report a compact set of metrics aligned with the model's intended use. For blind full inference, we emphasize total test loss together with Roman numeral (RN) accuracy, while also tracking component-level accuracies and structural F1 scores where available. Following~\cite{NapolesLopez2021_AugmentedNetRomanNumeral}, RN quality is primarily summarized with a Chord Symbol Recall (CSR)-style Roman-numeral score. Because several recent models also report per-note predictions~\cite{karystinaios2023chordgnn,kernfeld2025analysisgnn,sailor2024rnbert}, we additionally inspect note-level RN accuracy and observe no substantial difference in the qualitative conclusions on our validation runs. We therefore use Roman-numeral accuracy as the main harmonic summary and report component metrics separately for diagnostic detail.

For Table~\ref{tab:model_overview}, rows with citations report previously published baselines when directly comparable values are available, whereas rows marked \textit{retrain} denote our own reruns under the shared Dilemmadata schema using the RNBert (unconditioned) setting. Missing entries indicate that a prior model did not report that metric. This separation makes clear whether improvements come from stronger blind prediction or from better edit-conditioned completion.

Comparison follows a role-based protocol rather than a single leaderboard. The full \ours serves as the default blind-analysis baseline; the LoRA continuation is evaluated as a parameter-efficient continuation of that baseline; and the masked model is evaluated only in its intended partial-label regime. For decoding ablations, deltas are computed against each run's own blind full-pass branch, allowing us to isolate the effect of a specific inference-time module without conflating it with checkpoint differences.
Unless otherwise stated, the default blind full-inference configuration uses frozen cached MusicBERT note embeddings, the hybrid graph model, PCGrad for multitask optimization, mean pooling for onset-level reduction, and no post-hoc decoding modules. The optimization table therefore varies only the conflict-handling strategy and optional LoRA adaptation, while the decoding ablation table evaluates inference-time changes against this same blind full-pass reference.

For masked completion, a specified known-label ratio is used as conditioning context, and evaluation is computed only on masked positions. Known labels are selected according to the masking procedure described above rather than from recorded analyst corrections. Each ratio in Table~\ref{tab:masked_ratios} is evaluated with five independently sampled masks. This controlled protocol tests whether the model can exploit partial context, but it does not reproduce the order or localization of edits made by analysts.

\begin{table}[t]
    \centering
    \begin{tabular}{lcc}
    \toprule
    Known-label ratio & \textbf{AugNet} & \textbf{DLC} \\
    \midrule
    5\%  & .577 $\pm$ .001 & .573 $\pm$ .000 \\
    15\% & .592 $\pm$ .001 & .589 $\pm$ .001 \\
    25\% & .610 $\pm$ .004 & .605 $\pm$ .001 \\
    50\% & .667 $\pm$ .002 & .656 $\pm$ .001 \\
    75\% & .749 $\pm$ .004 & .727 $\pm$ .002 \\
    85\% & .788 $\pm$ .003 & .761 $\pm$ .002 \\
    95\% & .831 $\pm$ .008 & .801 $\pm$ .002 \\
    \bottomrule
    \end{tabular}
    \caption{Masked model performance at different known-label ratios. The percentage denotes the proportion of labels given to the model as context. AugNet and DLC report RN accuracy on the AugmentedNet and Distant Listening test sets, respectively. Accuracy is computed only on masked positions. Every run is repeated 5 times.}
    \label{tab:masked_ratios}
\end{table}

\begin{table}[t]
\centering
\small
\begin{tabular}{l|ccc}
\textbf{Conflict handling} & \textbf{AugNet} & \textbf{DLC} & \textbf{Cadence}\\
\hline
\multicolumn{4}{c}{\textit{Frozen MusicBERT + GNN}} \\
\hline
None    & .551          & .540 & .513 \\
GradNorm & .382          & .322          & .369\\
PCGrad  & \textbf{.576} & \textbf{.578} & .529\\
CAGrad  & .550          & .544 & \textbf{.638}\\
\hline
\multicolumn{4}{c}{\textit{MusicBert LORA Finetuning + GNN}} \\
\hline
PCGrad & .5575 & .5385 & .506\\
\end{tabular}
\caption{Blind full-inference on RN accuracy for \ours under different multitask conflict-handling strategies, first with frozen MusicBERT and then with LoRA adaptation using the best-performing frozen conflict handler. \textit{AugNet} and \textit{DLC} each correspond to one of the Dilemmadata test-sets, while \textit{cadence} reports the cadence F1 score on the the DLC testset. All rows use blind full inference without post-hoc decoding modules.}
\label{tab:hybrid_conflict_lora}
\end{table}


\begin{table}[bt]
\centering
\small
\begin{tabular}{l|c|c}
\textbf{Name} & \textbf{$\Delta$Loss}$\downarrow$ & \textbf{$\Delta$RN$_{\text{AugNet}}$}$\uparrow$ \\
\hline
Post-hoc voter & +0.0006 & -0.0013  \\
Iterative refinement & +0.5310 & +0.0092 \\
Beam search & +0.6707 & -0.0343 \\
Component-first refinement & +0.1175 & -0.0381 \\
Component-first beam & +0.9513 & -0.1146 \\
Component-first beam + rescorer & +0.9509 & -0.1146  \\
\end{tabular}
\caption{Configuration studies and optional decoding modules. ``Post-hoc voter'' denotes the learned weighted onset aggregator compared against mean pooling. ``Iterative refinement'' uses iterative decoding built on the masked-refinement machinery. ``Beam search'' denotes the onset-level beam decoder. ``Component-first refinement'' anchors the component tasks \texttt{localkey}, \texttt{degree1}, \texttt{degree2}, \texttt{quality}, and \texttt{inversion}. ``Component-first beam'' is the corresponding component-state beam decoder, shown both without and with its learned rescorer. All deltas are computed against the blind full-pass branch logged within the same run. Positive deltas are worse for loss and better for accuracies.}
\label{tab:ablation_results}
\end{table}

\section{Results}\label{sec:results}

The strongest current result is not a single model, but the combination of a robust full-analysis checkpoint, a dedicated masked completion checkpoint, and a flexible post-hoc inference stack. The full checkpoint provides the best blind-analysis baseline, while the masked model enables context-aware completion with perfect consistency on known Roman-numeral labels in the evaluated regime.

Table~\ref{tab:model_overview} compares the hybrid model against prior and retrained baselines under blind full inference. On AugNet, the hybrid system attains the best degree, local-key, and Roman-numeral scores, while remaining competitive on quality and inversion; the only higher single-task values come from RNBert on quality and ChordGNN+Post on inversion. The gains are even clearer on DLC, where the hybrid model achieves the best degree, quality, inversion, local-key, Roman, and phrase scores, while AnalysisGNN remains strongest on cadence. Overall, these comparisons suggest that combining pretrained contextual representations with graph-based relational reasoning improves RN accuracy without sacrificing broader multi-task coverage.

Table~\ref{tab:masked_ratios} shows that masked completion improves monotonically as more labels are provided as context. On AugNet, RN accuracy on masked positions increases from $0.577 \pm 0.001$ at 5\% known labels to $0.831 \pm 0.008$ at 95\%; on DLC, the same trend rises from $0.573 \pm 0.000$ to $0.801 \pm 0.002$. The gains become especially pronounced beyond 50\% context, which is consistent with the intended analyst-in-the-loop scenario where partial annotations provide strong local harmonic constraints.The low variance across the five mask samples suggests stable performance under the masking procedure.

Table~\ref{tab:hybrid_conflict_lora} shows that explicit conflict handling is important for the blind full-inference model. In the frozen MusicBERT setting, PCGrad yields the best RN accuracy on both AugNet ($0.576$) and DLC ($0.578$), improving clearly over training without a conflict handler. CAGrad produces the strongest cadence F1 ($0.638$) but does not match PCGrad on RN accuracy, indicating a task-level trade-off rather than a uniform improvement. LoRA adaptation of MusicBERT does not improve over the frozen PCGrad configuration in the current setup, which supports our choice of frozen pretrained representations as the default backbone for the main experiments.


Table~\ref{tab:ablation_results} summarizes the optional inference-time and post-hoc modules. Under blind full inference they are neutral at best: the learned onset voter is flat, iterative refinement gains a fraction of a point of RN accuracy at a large loss penalty, and all beam-search variants, including the component-first redesign with a learned rescorer, degrade accuracy. We read this not as a failure of the modules themselves but as evidence that blind full inference is the wrong evaluation regime for them: these mechanisms are designed to operate under constrained, edit-conditioned decoding, and their proper evaluation lives in the masked-completion setting reported in Table~\ref{tab:masked_ratios} and discussed in Section~\ref{sec:discussion}.

\section{Discussion}\label{sec:discussion}

A central obstacle to making neural RN analysis useful in practice is structural rather than modelling-based: symbolic scores are presented to sequence, graph, and slice-based networks alike as sets of notes, and analysis tasks are trained and evaluated as one label per task per note. Analysts, however, do not reason about notes in isolation; they work with harmonic labels placed at change-points and scoped to beats, measures, or longer segments. \ours{} and related systems therefore produce information at a finer granularity than analytical labels are naturally written and read. Adressing this gap, rather than only improving note-level accuracy, is one of the main motivations of this work.

We address this gap on the output side with a multi-level aggregation layer over Roman-numeral components. Per-note task probabilities are combined into top-3 Roman-numeral candidates, each with an explicit likelihood, at four levels: per note, per unique onset, per beat, and per measure. Clicking a note in the Verovio interface (Figure~\ref{fig:interface}) surfaces candidate labels for each of these scopes; selecting a candidate recolours the underlying task values in green where they agree and in red where they contradict the chosen overall label. This turns a mass of per-note softmax values into the object an analyst actually reads: a Roman-numeral label at a harmonic change-point, with an auditable breakdown of the evidence for and against it.

The same interface also connects model output to constrained re-prediction. Updating a single field regenerates the surrounding context through the masked-completion checkpoint, which treats the edited value as a hard constraint and re-predicts dependent labels accordingly. Table~\ref{tab:masked_ratios} shows that masked accuracy rises monotonically with the known-label ratio and passes blind full inference before the 50\% mark. We interpret this as evidence that the model can make effective use of partial analytical context, which is a key requirement for interactive revision and completion scenarios. Note that the post-hoc decoders in Table~\ref{tab:ablation_results} are not primarily intended as improvements for blind prediction, but as machinery for constrained, edit-conditioned use.

This framing also highlights a feature of symbolic analysis that metric-driven evaluation tends to hide: some non-trivial passages admit multiple defensible analyses. Different analysts may legitimately choose different modulation points, different chord roots in ambiguous voicings, and different segmentations of the same music, so that binary accuracy against a single reference is at best a coarse proxy for practical utility. The goal of the present paper is therefore not to claim a completed user-centered validation of analyst benefit, but to establish the modeling and system components needed to support such workflows: strong blind prediction, controllable completion from partial labels, and an interface for inspecting and revising model suggestions. 

\section{Conclusion and Future Work}\label{sec:conclusion}

We presented a framework that recasts symbolic RN analysis as a prediction-to-analysis continuum: a hybrid MusicBERT\,+\,graph backbone that makes blind full-score inference competitive with strong published baselines on Dilemmadata, paired with an edit-conditioned masked checkpoint and a Verovio-based interface that turns per-note probabilities into top-$k$ Roman-numeral candidates at note, onset, beat, and measure level. Beyond strong blind prediction, the framework supports constrained completion from partial labels, and masked completion improves monotonically as more contextual labels are provided.

Taken together, these results suggest that automatic RN analysis can be developed not only as a benchmark prediction task, but also as a technical foundation for future interactive analysis tools. In this paper, we evaluate the modeling capabilities required for such use, including blind inference, partial-label completion, and interface-supported inspection and revision, rather than the end-user benefits of a complete analyst-facing workflow.

Future work will quantify the practical value of this setup through studies with musicologists, focusing on annotation speed, correction behavior, and perceived usefulness during real analytical work. We also plan to improve the inference-time control modules for edit-conditioned decoding, add uncertainty estimates, and evaluate transfer across broader repertoires.

\section{Acknowledgements}
This research has been supported by the Swiss National Science Foundation (SNSF) through the project "Towards a Unified Model of Musical Form: Bridging Music Theory, Digital Corpus Research, and Computation" (grant no. 10000183; 2024-2028) and by the European Research Council (ERC) under the EU's Horizon 2020 research \& innovation programme, grant agreement No. 101019375 (Whither Music?).

\bibliography{ISMIRtemplate}

%
%
%
%

\end{document}